\documentclass[aps,prl,amsfonts,amssymb,amsmath,twocolumn,floatfix]{revtex4}
 
\usepackage{graphicx}
\usepackage{amsmath}

\newcommand*{\vJ}{\vec{J}}

\newcommand*{\cN}{{\cal N}}
\newcommand*{\cNf}{{\cal N}_f}
\newcommand*{\Area}{{\cal A}}

\begin{document}

\title{Interplay of Magnetic and Superconducting Proximity Effects in FSF Trilayers}

\author{Tomas L\"ofwander}
\author{Thierry Champel}
\author{Johannes Durst}
\author{Matthias Eschrig}

\affiliation{Institut f\"ur Theoretische Festk\"orperphysik,
Universit\"at Karlsruhe, 76128 Karlsruhe, Germany}

\date{July 16, 2005}

%\pacs{74.45.+c,74.78.Fk,75.75.+a}

\begin{abstract}
  We present theoretical results on the interplay of magnetic and
  superconducting orders in diffusive
  ferromagnet-superconductor-ferromagnet trilayers. The induced
  triplet superconducting correlations throughout the trilayer lead to
  an induced spin magnetization. We include self-consistency of the
  order parameter in the superconducting layer at arbitrary
  temperatures, arbitrary interface transparency, and any relative
  orientation of the exchange fields in the two ferromagnets. We
  propose to use the torque on the trilayer in an external magnetic
  field as a probe of the presence of triplet correlations in the
  superconducting phase.
\end{abstract}

\maketitle

The importance of triplet pairing correlations in the interface region
between a singlet superconductor and a ferromagnet recently became the
focus of research in the field of spintronics
\cite{ber01,kad01,esc03,buz05}. In contrast to clean triplet $p$-wave
superconductors and superfluids, for diffusive materials $p$-wave
correlations are suppressed and triplet correlations have $s$-wave
orbital symmetry, but are odd in frequency \cite{ber01}. In the case
of a homogeneous magnetization of the ferromagnet, the spin projection
of the triplet correlations on the quantization axis of the exchange
field is zero. If, on the other hand, the distribution of the exchange
field in the ferromagnet is inhomogeneous in space, then under
suitable conditions \cite{cha05a} also triplet correlations with
non-zero spin projection (equal spin pairs) are induced
\cite{ber01,kad01,esc03}. Triplet pairing correlations induce in turn
a spin magnetization both in the ferromagnet and in the superconductor
\cite{kri02,ber04a}.

An important question is how to experimentally find good fingerprints
of the triplet superconducting correlations. Theoretical work has been
focused on calculations of $T_c$ \cite{rad91,fom03,cha05a}, the local
density of states (LDOS) \cite{faz99,hal01,buz05}, or to search for
unconventional Josephson couplings \cite{vol03,esc03}. Experimentally,
no smoking gun has been found although recently magnetization changes
were observed through neutron reflectometry on multilayers of ${\rm
YBa}_2{\rm Cu}_3{\rm O}_7$ and ${\rm La}_{2/3}{\rm Ca}_{1/3}{\rm
MnO}_3$ \cite{sta05}. Various other properties of FS heterostructures
that could be influenced by triplet superconducting correlations have
been measured, as for example a negative Josephson coupling ($\pi$
junctions) \cite{rya01}, and LDOS modulations \cite{kon01}.

In 
%this Letter
the following
we present results for the induced triplet
correlations and corresponding changes of the magnetization in a
ferromagnet-superconductor-ferromagnet (FSF) trilayer with arbitrary
misalignment of the exchange fields in the two F layers. We put
forward signatures of triplet correlations that can be measured
experimentally.
\begin{figure}[b]
  \includegraphics[width=0.8\columnwidth]{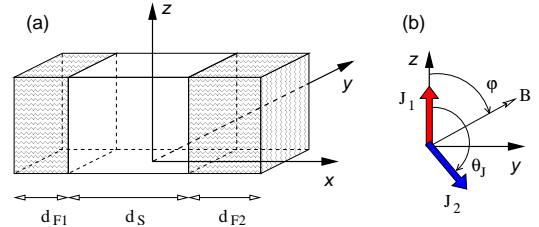}
  \caption{The trilayer consists of a superconductor of thickness
  $d_S$ and two ferromagnets of thicknesses $d_{F1}$ and $d_{F2}$. The
  exchange fields of the ferromagnets, $\vec J_1$ and $\vec J_2$, are
  confined to the $y$-$z$-plane, but are misaligned by an angle
  $\theta_J$. For torque measurements, a small magnetic field is
  applied in the $y$-$z$-plane at an angle $\varphi$.}
  \label{fig:world}
\end{figure}

The FSF trilayer we consider is sketched in Fig.~\ref{fig:world}. We
denote the layer thicknesses by $d_{F1}$, $d_S$, and $d_{F2}$
respectively. The $x$-axis is directed perpendicular to the layer
interfaces with the origin at the center of the superconductor. The
$z$-axis is aligned with the exchange field $\vJ_1$ in the left
ferromagnet. The angle between the exchange fields $\vJ_1$ and $\vJ_2$
is denoted $\theta_J$. We also assume translational invariance in the
$y$-$z$-plane.

We use the quasiclassical theory of superconductivity for diffusive
systems~\cite{usa70} that is formulated in terms of momentum averaged
Green functions. The Green function for the trilayer, $\hat
g(x,\epsilon_n)$, depends on the $x$-coordinate and on the Matsubara
frequency $\epsilon_n=(2n+1)\pi T$ ($n$ integer, $T$ temperature). In
standard notation (see e.g. \cite{cha05a,ale85}) the Green function is
a $4\times 4$ matrix in combined Nambu-Gor'kov (electron-hole) and
spin space
\begin{equation}\label{eq:green}
\hat g = \left(
\begin{array}{cc}
g_s + \vec g_t\cdot\vec\sigma & (f_s + \vec f_t\cdot\vec\sigma)i\sigma_y \\
(f_s^\ast + {\vec f}_t^\ast \cdot\vec\sigma^\ast )i\sigma_y  &
g_s^\ast + {\vec g}_t^\ast \cdot\vec\sigma^\ast
\\
\end{array}
\right),
\end{equation}
where $f_s$ and $\vec f_t$ are singlet and triplet pairing amplitudes,
$g_s$ and $\vec g_t$ are spin scalar and spin vector parts of the
diagonal Green function, and the vector
$\vec\sigma=(\sigma_x,\sigma_y,\sigma_z)$ is composed of Pauli spin
matrices. The Green function satisfy a Usadel-type equation
\cite{usa70,bul85}
\begin{equation}\label{eq:usadel}
\left[
i\epsilon_n\hat\tau_3 - \hat\Delta - \vJ\cdot\hat\sigma,\hat g
\right]
+\frac{D}{\pi}\partial_x(\hat g\partial_x\hat g) = \hat 0,
\end{equation}
where $D$ denotes the diffusion constant (equal to $D_F$ in F and to
$D_S$ in S), $[\hat a,\hat b]=\hat a \hat b-\hat b\hat a$, and
$\hat\sigma={\rm diag}(\vec\sigma,\vec\sigma^*)$.  The quantities
$\hat \tau_1$, $\hat \tau_2$, $\hat \tau_3$ and $\hat 1$ denote the
Pauli matrices and the unit matrix in electron-hole
space. Eq.~(\ref{eq:usadel}) is supplemented by a normalization
condition $\hat g^2=-\pi^2\hat 1$.

The superconducting singlet order-parameter $\hat \Delta$ is
determined by the pairing interaction, which we assume to be zero in
the ferromagnetic parts of the trilayer. By a proper gauge
transformation, $\hat\Delta$ can be chosen real, $\hat\Delta=\Delta
i\sigma_y\hat\tau_1$, where $\Delta$ satisfies the gap equation
\begin{equation}\label{eq:gap}
\Delta(x) \ln \left(\frac{T}{T_{c0}}\right) = \pi T \sum_{n} \left[
\frac{f_s(x,\epsilon_n)}{\pi} - \frac{\Delta(x)}{\epsilon_n} \right].
\end{equation}
The pairing interaction and the frequency-sum cut-off have been
eliminated as usual in favor of the bulk superconductor critical
temperature $T_{c0}$. The trilayer critical temperature is lower
$T_c<T_{c0}$, but we use $T_{c0}$ as a parameter independent energy
scale in our problem.

The interface boundary conditions have been formulated by Nazarov
\cite{naz99} and consist of two equations.  The first is the condition
of current conservation through the interface
$
\sigma_{F } \hat g_{F } \partial_x \hat g_{F } =
\sigma_{S } \hat g_{S } \partial_x \hat g_{S }
$,
where $\hat g_{F }$ and $\hat g_{S }$ denote the values of the Green
function at the F and S sides of the interface, respectively. The
normal state conductivities on either side are related to the
corresponding diffusion constants $D$ and densities of states at the
Fermi level $\cN_{f} $ by $\sigma =2\cN_{f} e^2 D $, where $e$ is the
electron charge. Note that $\cNf$ in the up and down spin bands are to
quasiclassical accuracy equal in the weak ferromagnet regime $T_c <
|\vJ| \ll E_f$, where $E_f$ is the Fermi energy. The second boundary
condition is written as \cite{note}
\begin{equation}\label{eq:bc}
\sigma_{F } \hat g_{F } \partial_x \hat g_{F } = \pm
\frac{1 }{\Area R_b} \frac{2\pi^2\left[\hat g_{F },\hat g_{S }\right]}{
2\pi^2(2-{\cal T})-{\cal T}\left\{\hat g_{F},\hat g_{S }\right\}},
\end{equation}
where $R_b$ is the boundary resistance, $\Area$ is the junction area,
and ${\cal T}$ is the junction transparency.  The sign $+$($-$) refers
to the left (right) interface.  For simplicity we use only two
parameters: ${\cal T}$ and $r_b=\Area R_b \sigma_F/(2\pi^2\xi_S)$,
where $\xi_S=\sqrt{D_S/2\pi T_{c0}}$ is the coherence length in S. At
the outer F surfaces the boundary conditions are $\partial_x \hat g_{F
}=0$.

We have solved Eqs.~(\ref{eq:usadel})-(\ref{eq:bc}) self-consistently
with the help of a parametrization of the Usadel Green function in
terms of Riccati amplitudes, as described in Ref.~\cite{esc04}. This
method allows us to treat spatially inhomogeneous (including
non-collinear) spin magnetizations.

\begin{figure}[t]
  \includegraphics[width=\columnwidth]{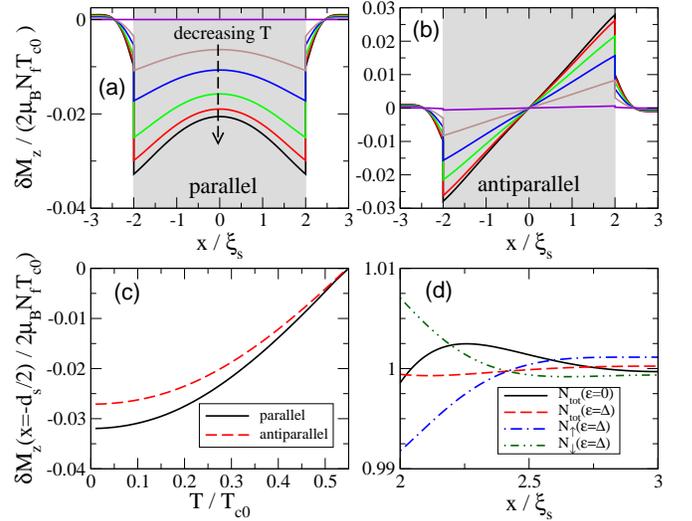}
  \caption{Spatial dependence of the induced magnetization at various
  $T\in[0.05,0.55]T_{c0}$ in steps of $0.1T_{c0}$ for (a) parallel
  ($\theta_J=0$, $\vec J_1 \uparrow \uparrow \vec J_2$) and (b)
  antiparallel ($\theta_J=\pi$, $\vec J_1 \uparrow \downarrow \vec
  J_2$) configurations.  The S region is shaded.  (c) Dependence of
  $\delta M_z$ on $T$ at the left interface (in S) for both
  configurations.  (d) Spatial dependence of the normalized LDOS in
  F$_2$ at $\epsilon=0$ and at $\epsilon=\Delta(x=0)=1.16T_{c0} $ for
  the parallel configuration.  Here, $d_{F1}=d_{F2}=\xi_S$,
  $d_S=4\xi_S$, $J_1=J_2=20T_{c0}$, $\sigma_F=\sigma_S$, $D_F=D_S$,
  and both interfaces have $r_b=0.1$ and ${\cal T}=1$.}
  \label{fig:magnetPAP}
\end{figure}
The magnetic order (exchange fields $\vJ_{1}$, $\vJ_{2}$) and the
superconducting order ($\Delta$) are spatially separated. However, due
to quantum mechanical leakage of the superconducting correlations
trough the interfaces both the singlet $f_s$ and the triplet $\vec
f_t$ pair correlations are spread out through the whole structure.  As
a result also the magnetization extends through the whole system,
including the superconductor. The induced spin magnetization below
$T_c$ is defined as
\begin{equation}\label{eq:magnetization}
\delta \vec M(x) = 2\mu_B\cNf 
                   T \sum_{n} \vec g_t(x,\epsilon_n),
\end{equation}
where $\mu_B$ is the Bohr magneton. In Fig.~\ref{fig:magnetPAP} we
present the induced magnetization within the trilayer for parallel and
antiparallel orientations of the exchange fields. We see in (a)-(b)
that $\delta \vec M$ extends into the superconductor over the distance
$\sim\xi_S$, while decaying and oscillating on the magnetic scale
$\xi_J=\sqrt{D_F/J}$ in the ferromagnets. There is a large increase in
the magnitude of $\delta\vec M$ as the temperature is lowered well
below $T_c$, see Fig.~\ref{fig:magnetPAP}(c). The corresponding
spatial variations of the LDOS \cite{noteLDOS} are shown in
Fig.~\ref{fig:magnetPAP}(d). The LDOS is spin split as due to the
presence of triplet correlations. The order of magnitude of the LDOS
modulations (1\% effect) is in agreement with experiments
\cite{kon01}.

\begin{figure}[t]
  \includegraphics[width=\columnwidth]{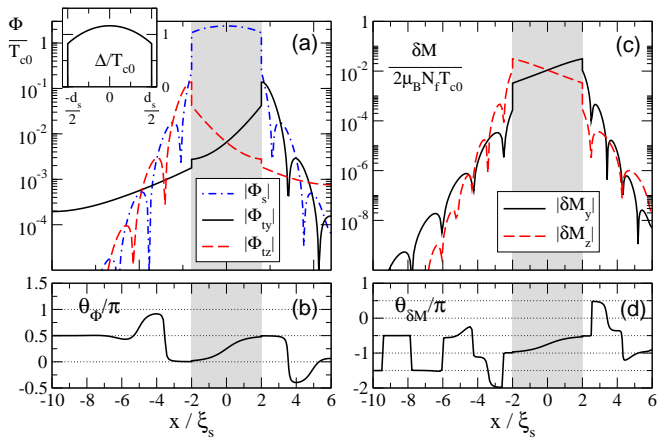}
  \caption{Spatial dependence (a) of the singlet ($\Phi_s$) and
  triplet ($\vec \Phi_t$) pairing amplitudes, and (c) of the induced
  magnetization $\delta \vec M$ for perpendicular configuration ($\vec
  {J}_1\perp \vec J_2$, $\theta_J=\pi/2$). Here, $d_{F1}=8\xi_S$,
  $d_{F2}=d_S=4\xi_S$, and $T=0.1T_{c0}$. The other parameters are as
  in Fig.~\ref{fig:magnetPAP}.  The S region is shaded.  (a) The
  singlet $\Phi_s(x)$ (dotted line) leaks into the F regions and
  oscillates and decays on a short length scale.  The triplets
  $\Phi_{ty}(x)$ (full line) and $\Phi_{tz}(x)$ (dashed line) are
  short-range or long-range depending on their respective projections
  on $\vec J_1$ or $\vec J_2$.  Inset: spatial dependence of
  $\Delta(x)$ in the S region.  (c) There is an asymmetry in the
  oscillation period between $\delta M_y$ (full line) and $\delta M_z$
  (dashed line) in both F regions.  (b) and (d): The angles
  $\theta_{\Phi }$ and $\theta_{\delta M}$ relative to the $z$-axis,
  that quantify the directions of $\vec \Phi_t$ and $\delta\vec M$ in
  the $y$-$z$-plane.
  \label{fig:magnet}}
\end{figure}
The singlet order parameter is suppressed at the interface to the
ferromagnets. At the same time, the singlet correlations leak into the
ferromagnets. For parallel or antiparallel orientations of the
exchange fields, triplet pairing correlations with zero spin
projections are induced. For other orientations equal spin pairs are
also induced. In Fig.~\ref{fig:magnet}(a)-(b) we show for
perpendicular orientation the spatial dependence of the order
parameter $\Delta(x)$ and of the correlation functions defined as
\cite{note2}
\begin{eqnarray}
\label{eq:order}
\Phi_s(x) &=& 2T \sum_{n>0} f_s(\epsilon_n,x),\\
\label{eq:order1}
\vec\Phi_t(x) &=&  T \sum_{n>0} \vec f_t(\epsilon_n,x).
\end{eqnarray}
Note that because $\vec f_t(-\epsilon_n,x)=-\vec f_t(\epsilon_n,x)$,
the total sum over all frequencies in Eq. (\ref{eq:order1}) would
vanish. Since we assume an energy-independent pairing interaction
there is no triplet order parameter. The singlet component is purely
real, while the triplet components are purely imaginary. The singlet
component and the triplet component with zero spin projection on the
local exchange field oscillate out of phase with respect to each
other, and decay fast in the ferromagnets on the magnetic length scale
$\xi_J$ \cite{buz82,rad91}.  As can be seen in
Fig.~\ref{fig:magnet}(a)-(b), there are long-range triplet pairing
correlations in both ferromagnetic layers. The component $\Phi_{ty}$
is decaying slowly in F$_1$ while $\Phi_{tz}$ is decaying slowly in
F$_2$. Relative to the local exchange fields, these components
describe the equal spin pairing correlations that decay on the
coherence length scale $\xi_T=\sqrt{D_F/2\pi T}$ but do not oscillate
\cite{ber01,kad01}. It should be noted, however, that these components
are already at the interfaces quite small since they are induced
non-locally in one ferromagnet by diffusion from the other through the
superconductor.

In Fig.~\ref{fig:magnet}(c)-(d) we show the spatial dependence of the
induced magnetization for perpendicular orientation of the exchange
fields. Again, $\delta\vec M$ is spread out over the large length
scale $\xi_S$ in the superconductor. But $\delta\vec M$ decays rapidly
in the ferromagnet on the scale $\xi_J$, because the spin-vector Green
function $\vec g_t$ is proportional to the fast decaying spin singlet
component $f_s$ through the relation \cite{cha05a}
$
g_s \vec g_t=f_s \vec f_t.
$
As a further consequence of this relation, the oscillation periods are
different for the two components of $\delta \vec M$. For example, in
F$_1$ the oscillation period of $\delta M_y$ is twice that of $\delta
M_z$. The reason is that $\delta M_y$ is determined by the product of
the oscillating $f_s$ and the monotonic $f_{ty}$ in
Fig.~\ref{fig:magnet}(a), while $\delta M_z$ is determined by the
product of the oscillating $f_s$ and the oscillating $f_{tz}$. Note
that deep in the ferromagnet (for $x<-5\xi_S$ in F$_1$), the
magnetization change is mainly due to the long-range triplet
correlations $f_{ty}$ and $\delta\vec M$ is therefore directed along
the $y$-axis. But the magnitude of $\delta\vec M$ is exponentially
small in this region. An observation of the two oscillation periods of
the two components $\delta M_z$ and $\delta M_y$, reflecting the
different behavior of the short-range oscillating $f_{tz}$ and the
long-range monotonic $f_{ty}$, would be a smoking gun for long-range
triplet components. Note that we have chosen a rather large value
$J=20T_{c0}$ in order to clearly separate the length scales in the
problem. Smaller values of $J$ might be more favorable in order to
experimentally resolve this effect.
\begin{figure}[t]
  \includegraphics[width=\columnwidth]{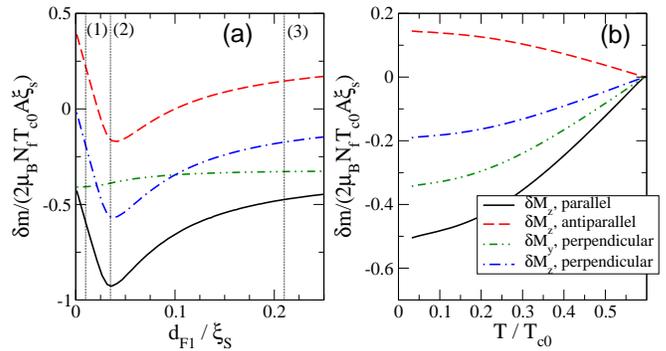}
  \caption{(a) Change in magnetic moment $\delta\vec m$ of a trilayer
  as function of the layer thickness $d_{F1}$, for $d_S=4\xi_S$,
  $d_{F2}=0.1\xi_S$, $T=0.1T_c$, and $J_1=J_2=20T_c$. The other
  parameters are as in Fig.~\ref{fig:magnetPAP}. (b) The temperature
  dependence of $\delta\vec m$ for parameters as in (a), but with
  $d_{F1}=0.2\xi_S$. The vertical lines in (a) are referred to in
  Fig.~\ref{fig:torque}.}
  \label{fig:moment}
\end{figure}

\begin{figure}[t]
  \includegraphics[width=8cm]{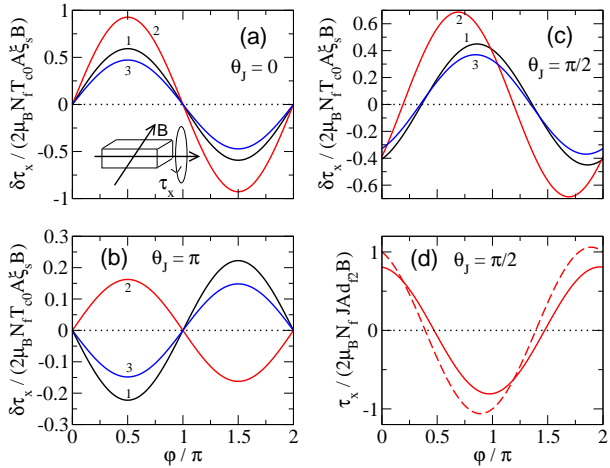}
  \caption{(a)-(c) The change in the torque relative to that in the
  normal state for parallel ($\theta_J=0$), antiparallel
  ($\theta_J=\pi$), and perpendicular ($\theta_J=\pi/2$)
  orientations. The parameters are the same as in
  Fig.~\ref{fig:moment} with the three numbered curves corresponding
  to the three layer thicknesses $d_{F1}=\{0.01,0.035,0.21\}\xi_S$
  (indicated by dashed vertical lines in Fig.~\ref{fig:moment}). In
  (d) we compare the total torque in the normal state (dashed line)
  with the one in the superconducting state (solid line from curve 2
  in (c)) within the itinerant ferromagnet model.}
  \label{fig:torque}
\end{figure}

Let us further address the issue of how to experimentally find
fingerprints of the triplet superconducting correlations.  We have
seen that the local magnetization change is rather small.  But it is
important to realize that the integrated effect,
\begin{equation}\label{eq:moment}
\delta\vec m = \Area \int dx \; \delta\vec M(x),
\end{equation}
can be large. To illustrate this, we show in Fig.~\ref{fig:moment} the
total magnetic moment of the trilayer, $\delta \vec m$, for various
orientations of the exchange fields, both as a function of $d_{F1}$
and of $T$.  We suggest to exploit the largeness of the integrated
effect by measuring the torque on the trilayer in a weak external
magnetic field. We consider a field $\vec B$ in the $y$-$z$-plane,
aligned at an angle $\varphi$ relative to the $z$-axis, as shown in
Fig.~\ref{fig:world}(b). Since the magnetic moment $\vec m$ of the
trilayer is also in the $y$-$z$-plane, the torque $\vec\tau=\vec m
\times \vec B$ is directed along the $x$-axis. We assume that the
external field is very weak so that the moment can be computed at
$B=0$. In the normal state there is a torque due to the magnetic
moment of the ferromagnetic layers. As the sample is cooled down
through $T_c$, the magnetic moment changes, and consequently the
torque is modified. This is illustrated for the parallel and
antiparallel exchange field orientations in
Fig.~\ref{fig:torque}(a)-(b). In addition, when the exchange fields
are not collinear, for example oriented perpendicular to each other as
in Fig.~\ref{fig:torque}(c), the $\varphi-$dependence changes
%. This
%occurs only if the structure is asymmetric 
for asymmetric structures (here the layer thicknesses are different,
$d_{F1}\neq d_{F2}$).  The shift of the maxima in $\delta
\tau_x(\varphi)$ when entering the superconducting state, as e.g. in
curve 2 in Fig.~\ref{fig:torque}(c), can be negative or positive
depending on the parameters of the trilayer.

In order to estimate the size of the change in the torque between the
normal and superconducting state, we use the itinerant ferromagnet
model. In this case the magnetization in the normal state is directly
related to the local exchange fields in the two layers as $\vec
M^N_{1/2}=2\mu_B\cNf \vec J_{1/2}$. The corresponding magnetic moments
are $\vec m^N_{1/2}=2\mu_B\cNf \vec J \Area d_{F1/2}$. The magnetic
moments in the normal and superconducting states are then $\vec
m^N=\vec m^N_1+\vec m^N_2$ and $\vec m^S=\vec m^N+\delta\vec m$,
respectively. In Fig.~\ref{fig:torque}(d) we show how the torque is
modified within this model.  In particular, the equilibrium
orientation ($\varphi$ for which $\vec m \parallel \vec B$) is
different in the normal and superconducting states and the maximum in
$\tau_x(\varphi)$ shifts to smaller $\varphi$ by $\sim $10\%.  The
magnitude of these effects increases when parameters are such that the
spin screening is more efficient (e.g. smaller $J$ or smaller
$d_{F1/2}$).

In summary, we have shown that superconducting triplet correlations in
FSF trilayers are much easier detected by measurements of global
properties of the sample rather than by local probes.  As an example
we have studied the induced magnetic moment in the superconducting
state.
The induced magnetic moment is a direct consequence of the presence of
triplet pairing correlations in the trilayer.
In an external magnetic field, the torque on the trilayer is
modified in the superconducting state. We have shown that this effect
can be more than 10\% for exchange fields as large as $|\vec J |\sim 20 T_{c0}$.
The proposed effect can be used to
experimentally determine the presence of triplet correlations.

\begin{acknowledgments}
  We acknowledge financial support from the EC under the spintronics
  network RTN2-2001-00440 (T.L.) and the 
  Deutsche Forschungsgemeinschaft within the Center for Functional Nanostructures
%DFG within the CFN
(T.C. and M.E.).
\end{acknowledgments}

 \vspace{-0.2cm}

%\bibliography{../../bibfiler/spintronics.bib}

\begin{thebibliography}{21}
\bibitem{note}
We assume here for simplicity
that all conductance channels contribute with equal weight.
\bibitem{noteLDOS}

The LDOS is calculated from the real energy version of
Eq.~(\ref{eq:usadel}). The spin resolved (along $\hat n$) LDOS is then
defined as
%
$ N_{\hat n} = -(\cNf/\pi) \left[g_s + \hat n \cdot \vec g_t \right],
$
%
while the total LDOS is the sum of spin up and spin down contributions.
\bibitem{note2}
These sums converge except for 
Eq. (\ref{eq:order}) in the S region, for which we used 
a frequency sum cut-off of 10 $T_{c0}$.
\end{thebibliography}

\end{document}